\documentclass{article}
\RequirePackage{natbib}
\usepackage{emulateapj,latexsym,apjfonts}
\newenvironment{inlinefigure}{%
\def\@captype{figure}%
\noindent\begin{minipage}{0.999\linewidth}\begin{center}}
{\end{center}\end{minipage}\smallskip}

\begin{document}

\submitted{}

\title{DIFFERENTIAL ROTATION IN NEUTRON STARS: MAGNETIC BRAKING AND VISCOUS DAMPING}

\author{Stuart L. Shapiro\altaffilmark{1}}

\affil{Department of Physics, University of Illinois at
        Urbana-Champaign, Urbana, Il 61801}

\altaffiltext{1}{Department of Astronomy and NCSA, University of Illinois at
        Urbana-Champaign, Urbana, Il 61801}

\begin{abstract}
Diffferentially rotating stars can support significantly more mass in 
equilibrium than nonrotating or uniformly rotating stars, 
according to general relativity. 
The remnant of a binary neutron star merger may give 
rise to such a ``hypermassive'' object. While such a star may be dynamically 
stable against gravitational collapse and bar formation, the radial 
stabilization due to differential rotation is likely to
be temporary. Magnetic braking and viscosity 
combine to drive the star to uniform rotation,
even if the seed magnetic field and the viscosity are 
small. This process inevitably leads 
to delayed collapse, which  will be accompanied by a 
delayed gravitational wave burst and, 
possibly, a gamma-ray burst.  We provide a simple, 
Newtonian, MHD calculation of the
braking of differential rotation by magnetic fields and viscosity. 
The star is idealized as a differentially rotating, 
infinite cylinder consisting of a homogeneous, incompressible
conducting gas.
We solve analytically the simplest case in which the gas has 
no viscosity and the star resides in an exterior vacuum.
We treat numerically cases in which the gas has 
internal viscosity and the star is embedded in an exterior, low-density,
conducting medium.  
Our evolution calculations are presented to 
stimulate more realistic MHD simulations in full $3+1$ general relativity. 
They serve to identify some of the key physical and numerical
parameters, scaling behavior and competing timescales that characterize 
this important process.

\end{abstract}

\section{Introduction}

%This is a $\mbox{\boldmath{$v$}}$.%
Neutron stars may form with 
appreciable differential rotation.
The merger of a binary neutron star system, for example,
will form a differentially rotating remnant, an outcome 
considerably enhanced by the liklihood that
the neutron stars are nearly irrotational before
coalescence (Bildsten and Cutler 1992; Kochanek 1992; 
Lai, Rasio \& Shapiro 1993). The formation
of a differentially rotating remnant following binary 
merger was first demonstrated in Newtonian hydrodynamic 
simulations (see e.g., Rasio and Shapiro 1992,1994,1999) 
and dramatically confirmed in fully relativistic simulations
(Shibata \& Uryu 2000), where the remnant's core is found to 
rotate appreciably faster than its envelope.   
Core collapse in a supernova may also result in a differentially rotating
neutron star, even if the rotation of the progenitor at the onset of collapse 
is only moderately rapid and almost uniform (see, eg, 
Zwerger \& M\"{u}ller 1997; Rampp,
M\"{u}ller \& Ruffert 1998, and references therein). Conservation of angular 
momentum during collapse causes the ratio $\beta = T/|W|$, where $T$ is 
the rotational kinetic energy and $W$ is the gravitational potential energy,
to grow like $R^{-1}$,
where $R$ is the equatorial radius. Thus $\beta$ can increase 
by several orders of magnitude during the implosion. 
Since uniformly rotating, compressible 
stars can only support very small values of $\beta$ in equilibrium 
without shedding mass at the equator (see, e.g., Tassoul 1978, 
Shapiro \& Teukolsky 1983), 
collapsed cores of rapidly rotating progenitors must acquire
some degree of differential rotation at birth as they settle into
equilibrium. Similar arguments hold for accretion-induced collapse of
white dwarfs to neutron stars.
 
Differentially rotating neutron stars can support significantly more rest mass
than their nonrotating or uniformly rotating counterparts. Baumgarte, Shapiro
\& Shibata (2000; hereafter BSS), have recently constructed relativistic 
equilibrium models of differentially rotating ``hypermassive''  
polytropes. These configurations 
have rest masses that exceed both 
the maximum rest mass of nonrotating spherical stars (the TOV limit)
and uniformly rotating stars at the mass-shedding limit (``supramassive
stars''), all with the same polytropic index.
BSS performed dynamical simulations in full general 
relativity to demonstrate that 
hypermassive stars can be constructed that are dynamically stable 
against radial collapse and nonradial bar formation 
(see also Shibata, Baumgarte \& Shapiro 2000).
The fully relativistic binary coalescence calculation of 
Shibata \& Uryu (2000) reveals
that such a dynamically stable hypermassive star can 
arise following binary merger. BSS have argued that
the dynamical stabilization of a hypermassive remnant by differential rotation
may lead to delayed collapse and a delayed gravitational wave burst.
The reason is that the stabilization due to differential rotation, 
although expected to last
for many dynamical timescales (i.e. many milliseconds), will ultimately be 
destroyed by magnetic braking and/or viscosity. These processes 
drive the star to uniform rotation, which cannot support the excess mass,
and will lead to catastrophic collapse, possibly accompanied by
some mass loss.

The purpose of this paper is to provide a time-dependent calculation 
of the damping of differential rotation  by magnetic braking and viscosity 
in a rapidly spinning star. The goal is to
construct an exact, albeit idealized, Newtonian model which readily reveals 
some of the important physical effects, together with 
their timescales and 
scaling behavior. By adopting a sufficiently simple, laminar 
model with a high degree of spatial symmetry, we are able to perform
a tractable analysis of the evolution from first principles and without
approximation.
Our motivation for this simplified approach is to help identify 
the minimum ingredients for a 
more realistic, future numerical calculation in full general relativity.
We wish to demonstrate explicitly why it is essential to include magnetic
fields in fully relativistic simulations of the late stages of 
coalescing binary neutron stars and
other dynamical scenarios where differential rotation naturally arises  
and significantly influences the structure and stability of the final 
configuration. The results have important implications for the
detection of gravitational wave and gamma-ray bursts (BSS; Spruit 1999a).

Our discussion is by no means the first, 
(see, e.g., Spitzer 1978
for discussion and references on magnetic braking of
angular momentum in interstellar clouds; see Spruit (1999b) and references
therein for a discussion of differential rotation and the evolution of
initially weak magnetic fields in stellar interiors), 
and surely will not be the last on this nontrivial topic,
especially given the simplified nature of our model. 
However, the importance of magnetic braking
has been pointed out only recently in the context of binary neutron star
coalescence, differential rotation of the remnant and the emission of 
gravitational waves following delayed collapse (BSS). 
Presumably the most effective
way to stimulate numerical relativists to include magnetic
fields in $3+1$ fully relativistic hydrodynamics codes now under
development is to provide a concrete calculation demonstrating 
their physical importance, which we do below.

In Section 2 we describe our basic model. In Section 3 we set out the
fundamental equations and put them in a convenient nondimensional form.
In Section 4 we summarize solutions for
three distinct cases, depending on whether the stellar exterior region
is a pure vacuum or filled with conducting plasma, and 
whether or not we include viscosity. In the first case (vacuum exterior, no
viscosity) our treatment is entirely analytic, while in the other 
two the analyses are numerical. In Section 5 we restore dimensions 
and evaluate our key results for the case of a differentially rotating remnant
of a binary neutron star merger. Here we summarize the implications of our
calculations to future observations and numerical simulations.  
Appendicies A and C summarize details
of our numerical calculation, while Appendix B provides a proof that
the lowest energy state for our adopted rotating stellar models is the state
of uniform rotation, for fixed angular momentum.

\section{Basic Model}

As discussed in BSS, 
differential rotation in a spinning star twists up a frozen-in, poloidal magnetic field, creating
a very strong toroidal field. This process will generate Alfv\'{e}n waves,
which can redistribute and even carry off angular momentum from a star on
timescales less than $\sim 100{\rm s}$ for poloidal fields  
greater than $\sim 10^{12}{\rm G}$. 
Shear viscosity also redistributes angular momentum. However, 
molecular viscosity
in neutron star matter operates on a typical timescale of 
$\sim 10^9{\rm s}$ in a
star of $\sim 10^9{\rm K}$, so it alone is much less 
effective in bringing the star 
into uniform rotation than magnetic braking, unless the initial magnetic 
field is particularly weak. Turbulent viscosity, if it arises, can act more
quickly. 

To track the evolution of differential rotation and follow the competition between
magnetic braking and viscous damping, we model a spinning star as an infinite, axisymmetric
cylinder. We take the gas interior to the star to be homogeneous in density and 
incompressible,
which is a crude but not unreasonable approximation for a neutron star governed by a stiff nuclear
equation state. We adopt cylindrical coordinates $(r,\phi,z)$, 
with the $z-$ axis aligned with the
rotation axis of the star. We take the magnetic field to have components only in the
$r$ and $\phi$ directions. We consider two cases: one in which the exterior region of the
star is pure vacuum and another in which it is a diffuse plasma at a constant density.
We treat the fluid everywhere to be perfectly conducting and allow for the presence of 
shear viscosity, which we take to be constant.  

A similar geometry
was adopted previously to study magnetic braking of a rotating,
cylindrical gas cloud immersed in the interstellar medium (Mouschouvias \& Paleologou 1979). 
However, in that analysis it proved sufficent to evolve the magnetic field only in the 
exterior, treating the cloud as a rigid rotator and ignoring viscosity altogether. 
Here we are specifically interested in the evolution of the field inside the star and 
learning exactly how the field and viscosity combine to alter the interior 
angular momentum profile of a differentially rotating configuration.

\section{Basic Equations}

The fundamental equations for our incompressible, perfectly conducting 
MHD fluid are the equation of continuity
\begin{equation}
\mathbf{\nabla}\cdot \mathbf{v}\ =\ 0,
\label{one}
\end{equation}
where $\mathbf{v}$ is the velocity, 
and the magnetic Navier-Stokes equation
\begin{eqnarray}
\label{two}
{\partial \mathbf{v} \over \partial t}\ 
&\!\!+\!\!&\ \left(\mathbf{v}\cdot \mathbf{\nabla}\right)
\mathbf{v}\ = \\ 
& & \!\! \ -\ {1\over\rho}\ \mathbf{\nabla} P\
-\ \mathbf{\nabla} \Phi\ 
+\ \nu \nabla^2 \mathbf{v}\ 
+\ {\left(\mathbf{\nabla}\times \mathbf{B}\right)
\times \mathbf{B} \over 4\pi\rho}, \nonumber
\end{eqnarray}
where $\rho$ is the density, 
$\nu $ is the viscosity, $P$ is the pressure, $\mathbf{B}$ is the
magnetic field and 
$\Phi$ is the Newtonian gravitational potential, which satisfies Poisson's
equation,
\begin{equation}
\nabla^2\Phi\ =\ 4\pi\rho.
\label{three}
\end{equation}
The magnetic field $\mathbf{B}$ 
satisfies Maxwell's constraint equation
\begin{equation}
\mathbf{\nabla}\cdot \mathbf{B}\ =\ 0,
\label{four}
\end{equation}
as well as the flux-freezing equation
\begin{equation}
{\partial \mathbf{B} \over \partial t}\ 
=\ \mathbf{\nabla}\times(\mathbf{v}\times \mathbf{B}).
\label{five}
\end{equation}

Solving equation (\ref{one}) for our cylindrical star gives the radial
component of the velocity as $v_r = R v_0(t)/r $ where R is the stellar radius.
Demanding the flow be regular at the origin requires that the function $v_0(t)$
vanish identically, implying that only rotational motion can take place, with an
azimuthal velocity given by $v_\phi = r \Omega(t,r)$. Here $\Omega$ is the
angular velocity.
Solving equation (\ref{four}) for the magnetic field 
together with the flux-freezing
condition (\ref{five}) requires that
the radial component of the magnetic field be independent of time and
given by
\begin{equation}
B_r = {B_0 R \over r}, ~~~t \geq 0,
\label{six}
\end{equation}
where $B_0$ is the value of the field at the stellar surface at $t=0$. 
Although the magnetic field given by 
equation (\ref{six}) exhibits a static line
singularity along the axis at $r=0$, it does not drive singular behavior in the
fluid velocity or nonradial magnetic field. As these quantities, which are
the main focus here, remain finite 
and evolve in a physically reasonable fashion, the line singularity poses no
difficulty and requires no special treatment.

Evaluating the azimuthal components of 
equations (\ref{two}) and (\ref{five}), the
system then reduces to two coupled partial differential equations 
for $\Omega$ and $B_{\phi}$,
\begin{equation}
{\partial\Omega \over \partial t}\ =
\ {RB_0 \over 4\pi\rho r^3} {\partial \over \partial r}
\left (rB_{\phi}\right)\ +\ {\nu \over r^3} {\partial \over \partial r}
\left (r^3 {\partial\Omega \over \partial r}\right),
\label{seven}
\end{equation}
and 
\begin{equation}
{\partial B_{\phi} \over \partial t}\ =\ B_0 R
{\partial\Omega \over \partial r}.
\label{eight}
\end{equation}

Equations (\ref{seven}) and (\ref{eight}) are the main dynamical equations to be solved.
Once $\Omega$ and $B_{\phi}$ are determined, the radial component of the Navier-
Stokes equation,
\begin{equation}
{\partial P \over \partial r}\ =\ \rho{v^2_{\phi} \over r}\
-\ \rho {\partial\Phi \over \partial r}\
-\ {1 \over 4\pi} \left({1 \over 2} {\partial B^2_{\phi} \over \partial r}\
+\ {B^2_{\phi} \over r}\right),
\label{nine}
\end{equation}
can be solved for the pressure as a function of $r$ and $t$ by a simple integration
over $r$. Though straightforward, obtaining the resulting pressure, which is the 
profile necessary to enforce hydrostatic equilibrium in the star in the radial 
direction at all times, is not required to obtain the evolution of the rotation
profile or magnetic field and is therefore of little interest here.
Similarly, the equation of heat transfer is not needed in solving the problem
of incompressible flow, as we are not interested here in the temperature
distribution. 

\subsection{Initial Conditions and Boundary Values}

We assume that no azimuthal component of the field $B_\phi$ 
is present initially,
\begin{equation}
B_{\phi}(0,r) = 0, \,\,\,\,  r \geq 0,
\label{ten}
\end{equation}
recalling that $B_r$ is always given by equation (\ref{six}). We are thus
interested in the situation where the azimuthal field is created entirely
by the differential rotation of the fluid, which bends the frozen-in,
initially pure radial field lines in the azimuthal direction.

We solve equation (\ref{seven}) for $\Omega$ subject to the condition of
regularity at the origin,
\begin{equation}
{\partial\Omega(t,0) \over \partial r}\ =\ 0, ~~~ t \geq 0,
\label{eleven}
\end{equation}
which, by equations (\ref{eight}) and (\ref{ten}), implies
\begin{equation}
{\partial B_{\phi}(t,0) \over \partial t}\ =\ 0\ =\ B_{\phi}(t,0),
~~~  t \geq 0.
\label{twelve}
\end{equation}

The boundary conditions at the stellar surface depend on the physical
situation being investigated. In the case of an exterior vacuum,
no azimuthal magnetic field can be carried into the region outside the
star. This fact, together with equation (\ref{eight}), implies
\begin{equation}
B_{\phi}(t,R)\ =\ 0\ =\ {\partial\Omega(t,R) \over \partial r},
~~~ t \geq 0 ~~~~ {\rm (vacuum\ exterior).}
\label{thirteen}
\end{equation}
In the case where the exterior region of the star contains a conducting
homogeneous fluid at density $\rho_{\rm ex}$, part of the azimuthal field is 
transmitted across the stellar surface in the form of an outgoing 
Alfv\'{e}n wave, 
and part is reflected via an ingoing Alfv\'{e}n wave. We construct an 
appropriate wave boundary condition accounting for partial 
transmission and reflection at the interface between the interior star and
the exterior plasma according to 
\begin{equation}
\hspace{-0.08cm}{\partial B_{\phi}(t,R) \over \partial t}\
+\ {\partial B_{\phi}(t,R) \over \partial r} \ v_{\rm A}\ {(1+{\cal R})
\over (1-{\cal R})}\ = \ 0 ~~ {\rm (plasma\ exterior),}
\label{fourteen}
\end{equation}
where $\mathcal{R}$ is the 
reflection coefficient for the azimuthal field amplitude,
\begin{equation}
{\cal R}\ =\ {(\rho_{\rm ex}/\rho)^{1/2}-1 \over (\rho_{\rm ex}/\rho)^{1/2}+1},
\label{fifteen}
\end{equation}
and $v_{\rm A} = B_0/(4 \pi \rho)^{1/2}$ is the Alfv\'{e}n 
speed just inside the stellar surface (see Appendix A). 
The corresponding boundary condition for $\Omega$ is
\begin{equation}
{\partial\Omega(t,R) \over \partial r} \
+ \ {\partial\Omega(t,R) \over \partial t} {1 \over v_{\rm A}}
{(1+{\cal R}) \over (1-{\cal R})} \ =\ 0 ~~~~ {\rm (plasma\ exterior).}
\label{sixteen}
\end{equation}
Note that when $\rho_{\rm ex} = 0$, we have ${\mathcal{R}}= -1$, so 
equations (\ref{fourteen}) and (\ref{sixteen}) 
recover the vacuum exterior boundary conditions (\ref{thirteen}),
recalling equation (\ref{ten}). By contrast, when $\rho_{\rm ex} = \rho$,
and the exterior plasma is a smooth continuation of the interior star,
we have ${\mathcal{R}} = 0$, and equations (\ref{fourteen}) and
(\ref{sixteen}) impose standard outgoing (Alfv\'{e}n) wave boundary 
conditions on these transverse field and velocity components.

The entire evolution of the system is driven by differential rotation.
We take the initial rotation profile of the star to have the following 
nonuniform, momentarily stationary form:
\begin{equation}
\Omega(0,r)\ =\ {1 \over 2} \Omega_0 [1+{\rm cos}(\pi r^2/R^2)],
~~~ 0 \leq r \leq R,
\label{seventeen}
\end{equation}
where
\begin{equation}
{\partial\Omega(0,r) \over \partial t}\ =\ 0, ~~~ 0 \leq r \leq R.
\label{eighteen}
\end{equation}
We note that our adopted initial angular velocity profile is 
consistent with the boundary conditions at the center and surface of
the star (see equations (\ref{eleven}) and (\ref{thirteen})), as
required.

\subsection{Conserved Energy and Angular Momentum}

The MHD equations (\ref{seven}) and (\ref{eight}) admit two nontrivial 
integrals of the motion, one expressing conservation of energy and the other
conservation of angular momentum of the star. Energy conservation is given by the
integral
\begin{equation}
E_{\rm rot}(0)\ =\ E_{\rm rot}(t)\ +\ E_{\rm mag}(t)\
+\ E_{\rm vis}(t)\ +\ E_{\rm Poyn}(t),
\label{nineteen}
\end{equation}
where $E_{\rm rot}(t)$ is the rotational kinetic energy of the matter, $E_{\rm mag}(t)$ is the 
azimuthal magnetic
energy, $E_{\rm vis}(t)$ is the  internal (thermal) energy generated by viscous
dissipation and $E_{\rm Poyn}(t)$ is the energy carried off at 
the stellar surface
by the Poynting vector $\bf{S} = c \bf{E} \times \bf{B} /4 \pi$,
where $\bf{E} = -\bf{v} \times \bf{B}/c$: 
\begin{eqnarray}
\label{tw}
E_{\rm rot}(t) = \int d^3x (\rho\Omega^2(t,r)r^2/2) &\!\!,\!\!&
E_{\rm mag}(t) = \int d^3x (B^2_{\phi}(t,r)/8\pi) \nonumber \\
E_{\rm vis}(t) =\int^t_0 dt \dot{{\cal E}}_{\rm vis}(t) &\!\!,\!\!&
E_{\rm Poyn}(t) = \int^t_0 dt \dot{{\cal E}}_{\rm Poyn}(t).
\end{eqnarray}
The rates of viscous dissipation
$\dot{\mathcal E}_{\rm vis}(t)$ 
and Poynting energy loss
$\dot{\mathcal E}_{\rm Poyn}(t)$ appearing in equation (\ref{tw}) 
are given by 
\begin{eqnarray}
\dot{{\cal E}}_{\rm vis}(t)\ &  = & \ \int d^3x \left[\nu\rho \left({\partial\Omega \over
\partial r}\right)^2r^2\right], \\
~~ \dot{{\cal E}}_{\rm Poyn}(t)\ & = & \
- \left({B_\phi B_r\Omega R \over 4\pi}\right)_{r=R} {\cal A}. \nonumber
\label{twone}
\end{eqnarray}
The volume elements in equations (\ref{tw}) and (\ref{twone}) are
evaluated as $d^3 x = 2 \pi L r dr$, with the volume 
integration performed throughout the interior of the cylindrical star 
over a (arbitrary) height $L$ in the
$z-$direction. The Poynting flux is measured at the surface of this 
region, whose area is ${\mathcal A} = 2 \pi L R$.
Dividing all quantities by $L$ then gives energies per unit
length. 

Conservation of  angular momentum is expressed as
\footnote{In deriving both equations (\ref{nineteen}) and (\ref{twtwo}) we
have set the viscous surface terms proportional to $\nu d \Omega/dr$ equal to
zero, which is appropriate for all of the cases treated below.} 
\begin{equation}
J_{\rm rot}(0)\ =\ J_{\rm rot}(t)\ +\ J_{\rm mag}(t),
\label{twtwo}
\end{equation}
where $J_{\rm rot}$ is the rotational angular momentum of the matter and $J_{\rm mag}$
is the integrated torque exerted by the Maxwell stress at the surface,
\begin{equation}
J_{\rm rot}(t)\ =\ \int d^3x (\rho\Omega(t,r)r^2),~~~ J_{\rm mag}(t)\ 
=\ \int^t_o dt {\cal N}.
\label{twthree}
\end{equation}
The torque $\mathcal N$ is given by
\begin{equation}
{\cal N}\ =\ - \left({B_rB_{\phi}R \over 4\pi}\right)_{r=R} {\cal A}.
\label{twfour}
\end{equation}
We note that, consistent with the nonrelativistic MHD approximation, the electric field
energy $E^2/8\pi$ is not included in equation (\ref{nineteen}) and the angular momentum
of the electromagnetic field $ {S}_{\phi}/c^2$ is not included in 
equation (\ref{twtwo}) (Landau, Lifshitz \& Pitaevskii 1984).

The motivation for the monitoring the conservation equations during 
the evolution is twofold: physically, evaluating the individual terms enables us
to track how the initial rotational energy and angular momentum in the fluid 
are transformed, dissipated and/or transported away; 
computationally, monitoring how well the conservation equations are 
satisfied provides a check on the numerical integration scheme.

\subsection{Nondimensional Formulation} 

We introduce nondimensional variables to facilitate the integration and to help
identify the scaling behavior of our solution with respect to our arbitrary choices of 
input parameters. We define nondimensional quantities according to
\begin{eqnarray}
\label{twfive}
& & \hat{r} = r/R,\;\;\; \hat{t} = 2t/(R/v_{\rm A}),\;\; \;
\hat{\nu} = 4\nu/(v_{\rm A}R)
\\ \nonumber
& & \hat{\Omega} = \Omega/\Omega_0,\;\;\; \hat{B}_\phi =
B_\phi/\left ((4\pi\rho)^{1/2}(\Omega_0R) \right) \\
& & \hat{E} = E/\left( (\pi R^2L\rho)(\Omega^2_0R^2/2) \right),\;\;\; \hat{J} =
J/\left( (\pi R^2L\rho)(\Omega_0R^2) \right). \nonumber
\end{eqnarray}
We work with nondimensional variables in all subsequent equations, but for simplicity we omit
placing carets (~$\hat {}$~) on the variables.
In nondimensional variables, the coupled evolution equations (\ref{seven}) and (\ref{eight}) become
\begin{equation}
{\partial\Omega \over \partial t} =  {1 \over r^2}
{\partial \over \partial r^2}(rB_\phi)\ +\ \nu {\partial \over \partial r^4}
 \left(r^4 {\partial\Omega \over \partial r^2}\right),
\label{twsix}
\end{equation}
and
\begin{equation}
{\partial B_\phi \over \partial t} =  r {\partial\Omega \over \partial r^2}.
\label{twseven}
\end{equation}
When expressed in nondimensional units, the initial value equations and 
boundary conditions take on the same appearance as in
equations (\ref{ten}) -- (\ref{eighteen}), provided
we set $v_{\rm A} = 1 = R$ wherever these quantities appear.
Multiplying equation (\ref{twsix}) by $\Omega r^2 d r^2$, integrating over 
the radius of the 
star and making use of use equation (\ref{twseven}) yields  
energy conservation equation (\ref{nineteen}), 
where the  nondimensional energy variables now appearing in that equation are
\begin{eqnarray}
\label{tweight}
& E_{\rm rot}(t)=\int^1_0 dr^2 \Omega^2(t,r)r^2,\;\;\;
E_{\rm mag}(t)\ =\ \int^1_0 dr^2 B^2_\phi ~, & \\
& E_{\rm vis}(t)= \nu \int^t_0 dt \left[\int^1_0 dr^2
\left({\partial\Omega \over \partial r^2}\right)^2 r^4\right], & 
\nonumber \\ 
& E_{\rm Poyn}(t)\ =\ - 2 \int^t_0 dt(B_\phi\Omega)_{r=1} ~. & \nonumber
\end{eqnarray}
Multiplying equation (\ref{twseven}) by $r^2 d r^2$ and integrating 
over the star $\rm {yields}^2$
angular momentum conservation equation (\ref{twtwo}), where the nondimensional
angular momentum terms are given by 
\begin{equation}
J_{\rm rot}(t)\ =\ \int^1_0 dr^2 \Omega(t,r)r^2 ~,
\ \ \ J_{\rm mag}(t)\ =\ -\int^t_0 dt (B_\phi)_{r=1}.
\label{twnine}
\end{equation}

The coupled system (\ref{twsix}) and (\ref{twseven}) is purely hyperbolic in the
absence of viscosity and leads to a simple wave equation. The inclusion of viscosity
gives the system a mixed hyperbolic-parabolic form, since viscosity contributes 
diffusive damping to the evolution.

\section{Solutions} 

We solve the evolution equations for three different cases. In case A, we
treat a viscous-free star with a vacuum exterior. In case B, we consider
the effect of viscosity. In case C, we again ignore viscosity but assume that
the exterior region consists of a perfectly conducting plasma of lower
density than the central star. By examing these three cases separately 
we are able to distinguish the different physical processes by which 
the rotation profile in the star is altered.

\subsection{Case A: ~~$\nu = 0$ with vacuum exterior}

Here we treat the simplest case in which the fluid is assumed to have no viscosity and
the region exterior to the star is a pure vacuum. 
Setting $\nu = 0$ we take the partial derivative of equation (\ref{twsix}) with respect to time
and use equation (\ref{twseven}) to eliminate $B_{\phi}$, thereby obtaining 
a simple wave equation for $\Omega$,
\begin{equation}
{\partial^2\Omega \over \partial t^2}\ =\ {1 \over r^2} 
{\partial \over \partial r^2} \left(r^2{\partial\Omega \over \partial r^2}\right).
\label{th}
\end{equation}
This equation must be solved subject to the boundary conditions (\ref{eleven}) 
and (\ref{thirteen})  
\begin{equation}
{\partial\Omega(t,0) \over \partial r}\ =\ 0\ =\ {\partial\Omega(t,1) \over \partial r},
\label{thone}
\end{equation}
and initial conditions (\ref{seventeen}) and (\ref{eighteen})
\begin{equation}
\Omega(0,r)\ =\ {1 \over 2} [1 + {\rm cos}(\pi r^2)],
~~{\partial\Omega(0,r) \over \partial t}\ =\ 0
\label{thtwo}
\end{equation}
%({\it cf.} Mouschovias and Paleologou 1979, who obtain a similar 
%equation, but examine it for a different radial domain of 
%dependence subject to different boundary and initial conditions.)
We recognize equation (\ref{th}) as $\Box \Omega = 0$ expressed in cylindrical coordinates, 
with cylindrical radial coordinate $\xi = r^2$ and no dependence on $\phi$ or $z$. In these 
coordinates, the equation is identical to the one that would arise when determining
the amplitude of the oscillations of a circular, axisymmetric drumhead, 
{\it unclamped} at the outer 
edge (see, e.g, Mathews \& Walker 1970 for an
analysis of a circular drumhead that is {\it clamped} at the outer edge). We  
solve the equation analytically by separation of variables, obtaining an expansion
over Bessel functions according to
\begin{equation}
\Omega(t,r)\ =\ \sum^\infty_{n=1}\ =\ A_nJ_0(k_nr^2){\rm cos} (k_nt),
\label{ththree}
\end{equation}
where
%\vspace{-0.45cm}
\begin{equation}
%\vspace{0.2cm}
A_n\ =\ {2\int^1_0 dr^2r^2 J_0(k_nr^2)\Omega(0,r) \over J^2_0(k_n)},
\label{thfour}
%\vspace{0.2cm}
\end{equation}
and where $\Omega(0,r)$ is set by equation (\ref{thtwo}). 
In the above expansion, the quantities $\{k_n\}, n = 1,2, \ldots$ are the zeroes of the derivative
of $J_0 (x)$, i.e., they satisfy $d J_0(k_n)/dx = - J_1(k_n) = 0$ and
are tabulated (Abramowitz and Stegun (1972), Table 9.5, p. 409, remembering 
to add $k_1 = 0$). The azimuthal field is obtained by substituting equation (\ref{ththree}) into
equation (\ref{twseven}), integrating over time and using equation (\ref{ten})
to set the initial value. The result is
%\vspace{-0.45cm}
\begin{equation}
%\vspace{0.2cm}
B_\phi\ =\ -r \sum^\infty_{n=1}A_nJ_1(k_nr^2) {\rm sin} (k_n t).
\label{thfive}
%\vspace{0.2cm}
\end{equation}

We evaluate the coefficients $A_n$ up to $n=20$ by quadrature and the resulting standing wave
solutions for the angular velocity and azimuthal magnetic field are shown
in Figs.~1 -- 3.  In Fig.~1 we track the
evolution of the various energies with time and observe how the total conserved energy oscillates
between  rotational kinetic energy of the fluid  and  energy of the azimuthal magnetic field.
The oscillations proceed forever without any dissipation or energy loss. Angular momentum,
which is carried entirely by the fluid, is also strictly conserved. We plot the angular
velocity profile at critical phases during an oscillation period ($P = 1.64$) in Fig.~2 and 
we plot the corresponding azimuthal field (a standing Alfv\'{e}n wave) in Fig.~3. 

The series solutions (\ref{ththree}) and (\ref{thfive}) converge rapidly: the first few
values of $A_n$ are $A_1 = 0.2974$, $A_2 = 0.7192$, $A_3 = -0.0198$ and $A_4 = 0.0044$ .
We already obtain a reasonable approximation by using the first two coefficients alone, together
with the quantities $k_1 =0$ and $k_2 = 3.8317$, to get 
\begin{equation}
%\vspace{-0.3cm}
\Omega(t,r) \approx 0.2974 + (0.7192)J_0(3.8317~r^2){\rm cos}(3.8317~t),
\label{thsix}
%\vspace{-0.3cm}
\end{equation}
and
%\vspace{-0.5cm} 
\begin{equation}
%\vspace{-0.3cm}
B_{\phi} \approx -(0.7192)rJ_1(3.8317~r^2){\rm sin}(3.8317~t).
%\vspace{-0.3cm}
\label{thseven}
\end{equation}
%%%%%%%%%%%%%%%%
%figure 1

\vspace{-2cm}
\begin{inlinefigure}
\centerline{\includegraphics[width=1.0\linewidth]{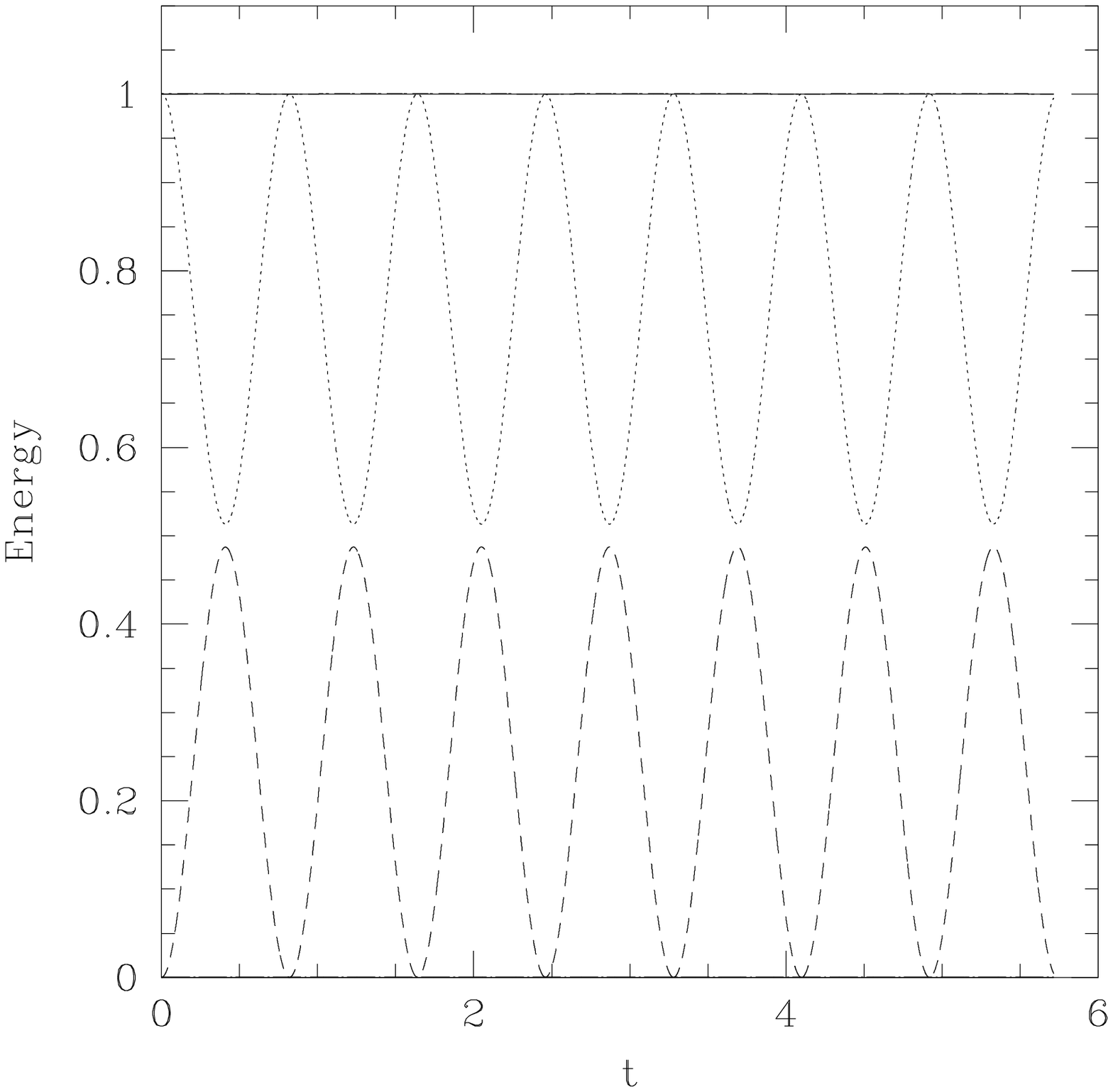}}
\vspace{-2cm}
\figcaption{Energy evolution for a differentially rotating star with zero viscosity and 
a vacuum exterior. The dotted
line shows $E_{\rm rot}$ and the dashed line shows $E_{\rm mag}$. 
All energies are normalized
to $E_{\rm rot}(0)$. The sum of the energies is conserved and remains
equal to the initial rotational energy $E_{\rm rot}(0)$, which is
plotted as the solid horizontal line at $\rm{Energy} = 1$. Time is in 
nondimensional
units.}
%\label{fig:analytics}}
\end{inlinefigure}
%%%%%%%%%%%%%%%%
%%%%%%%%%%%%%%%%
%\vspace{-2cm}
%figure 2
\begin{inlinefigure}
\vspace{-2cm}
\centerline{\includegraphics[width=1.0\linewidth]{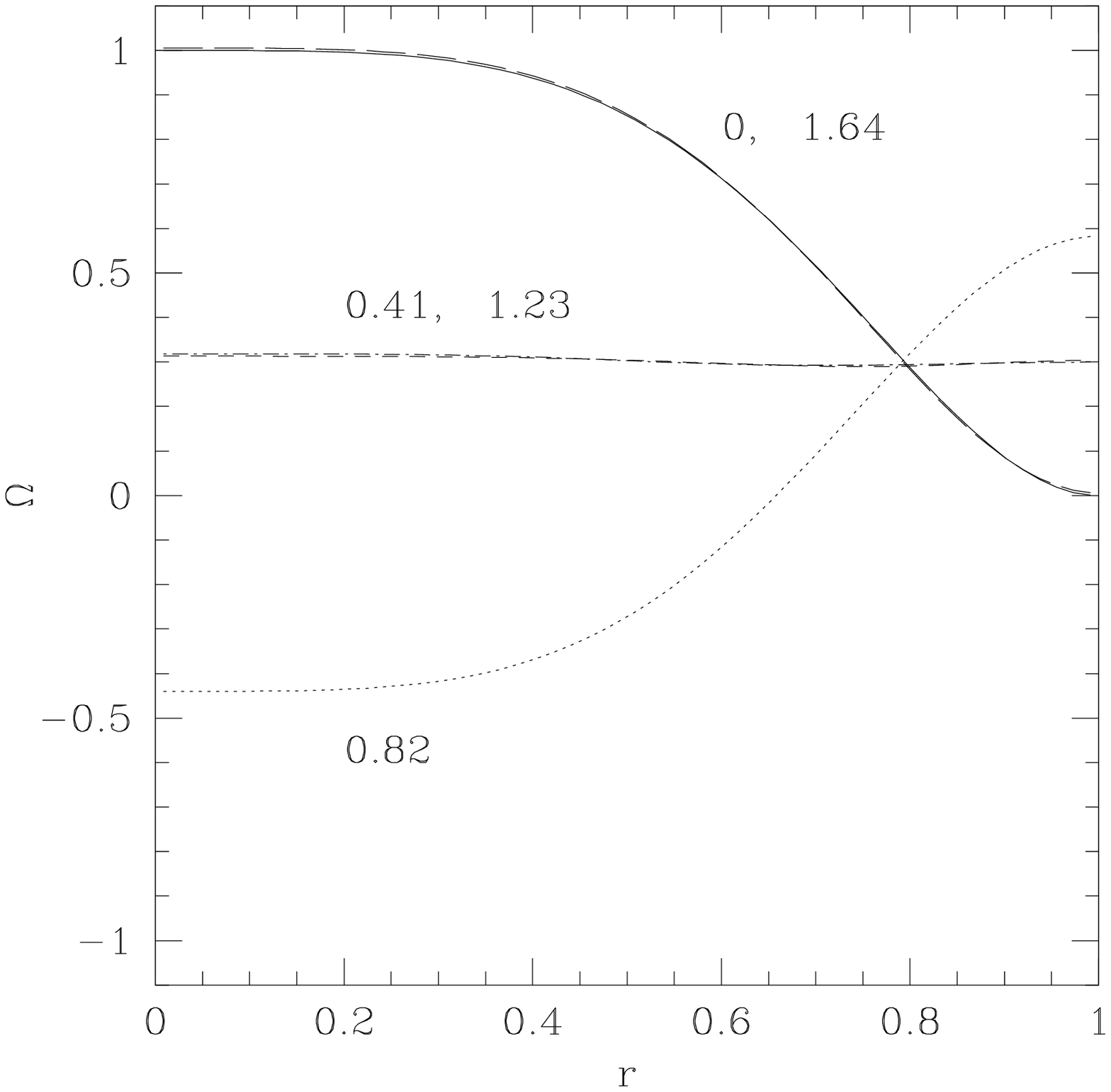}}
\vspace{-2cm}
\figcaption{The angular velocity profile at critical phases during the first oscillation cycle for
the differentially rotating star described in Fig.~1. Curves are labeled by the
value of time; the standing wave pattern oscillates with a period $ P = 1.64$. All quantities
are plotted in nondimensional units.}
%\label{fig:analytics}}
\end{inlinefigure}
%%%%%%%%%%%%%%%%
%%%%%%%%%%%%%%%%
%figure 3
\begin{inlinefigure}
\vspace{-2cm}
\centerline{\includegraphics[width=1.0\linewidth]{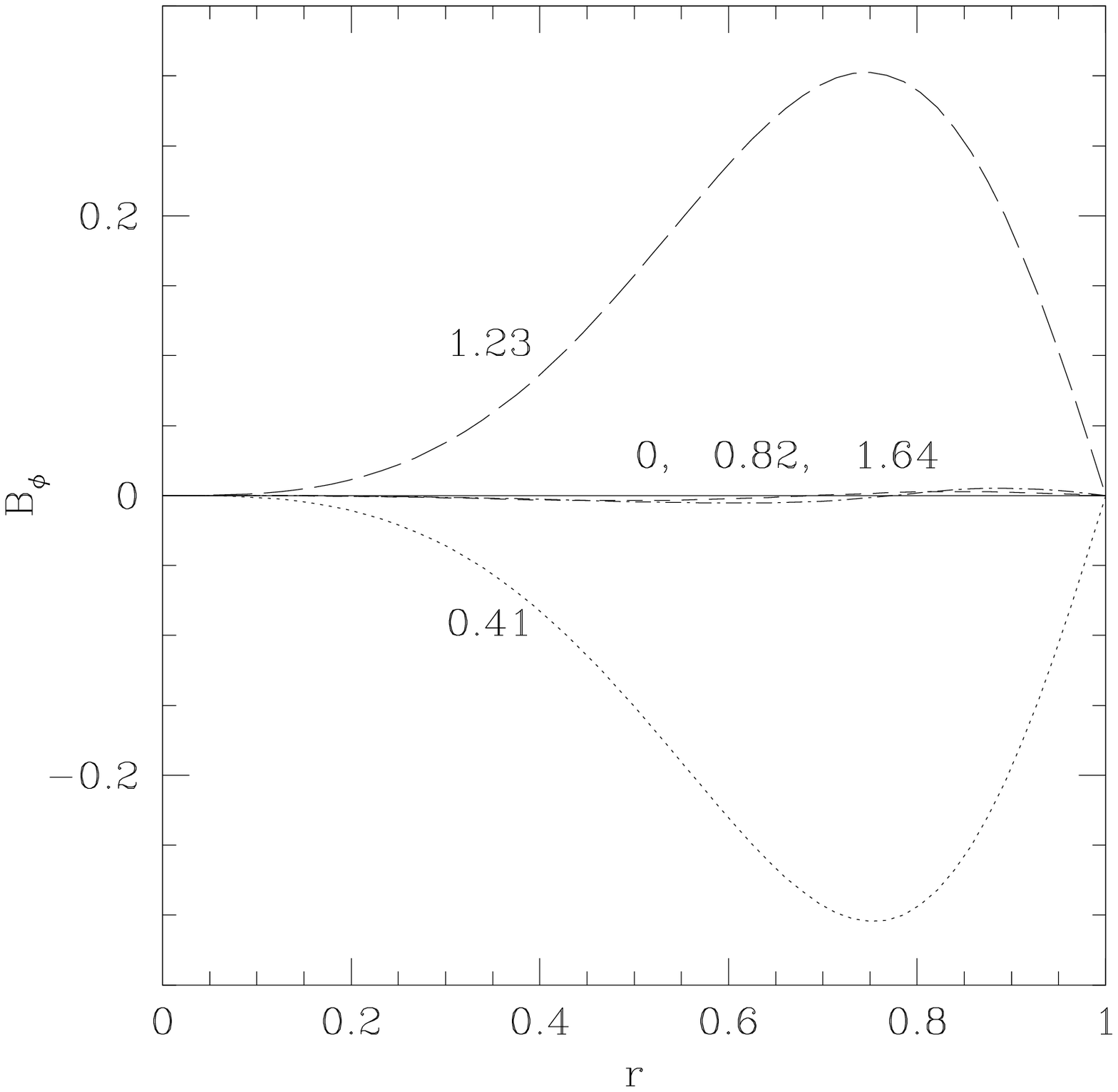}}
\vspace{-2cm}
\figcaption{The azimuthal magnetic field profile at critical phases during the first oscillation cycle for
the differentially rotating star described in Fig.~1. Curves are labeled by the
value of time; the standing Alfv\'{e}n wave pattern repeats with a period $ P = 1.64$. 
All quantities are plotted in nondimensional units.}
\vspace{0.5cm}
%\label{fig:analytics}}
\end{inlinefigure}
%%%%%%%%%%%%%%%%
From the above approximation we derive the oscillation period to be $P \approx 2 \pi / k_2 = 1.64$,
consistent with the behavior shown in Figs.~1 -- 3. From equation (\ref{thsix}) we
calculate $\Omega_{\rm unif}$, the value of the angular velocity at the phases
of uniform rotation, to be $\Omega_{\rm unif} = A_1 = 0.2974$,
consistent with the horizontal lines in Fig.~2. This value 
of $\Omega$ can also be found by invoking angular momentum conservation: 
setting $J_{\rm rot}(0)$, the total (initial) angular momentum of our differentially 
rotating star, obtained using equations (\ref{twnine}) and (\ref{thtwo}), to 
the angular momentum of a rigidly rotating configuration with the same
density profile, we find
\begin{equation}
\Omega_{\rm unif} = {{\pi}^2 -4 \over 2 {\pi}^2} = 0.2974
\label{theight}
\end{equation}
At these phases $B_{\phi}$ achieves its maximum amplitude.
The energy at the uniformly rotating phases is 
distributed more evenly between rotational kinetic and azimuthal magnetic field energy.
In our model, the rotational energy falls to a fraction 
$E_{\rm rot}/E_{\rm rot}(0) = 2({\pi}^2-4)^2/ ({\pi}^2 (3 {\pi}^2-16)) = 0.513$, 
while the magnetic energy increases to
$E_{\rm mag}/E_{\rm rot}(0) = 1 - E_{\rm rot}/E_{\rm rot}(0) = 0.487$ at
these phases. In the absence of dissipation, the star oscillates about the uniformly rotating
state indefinitely.

The presence of
internal energy dissipation (e.g., viscosity), however small, will drive a differentially rotating
star to a {\it permanent} state of uniform rotation, since uniform rotation is 
the lowest energy state at 
fixed angular momentum (see Appendix B for a proof). In this situation, the dissipated energy
ultimately goes into heat, and the azimuthal magnetic field decays away. We will 
study this behavior in Case B, below. However, even in the absence of dissipation, the presence
of a toroidal (here, radial) magnetic field will brake the differential rotation
on the time it would take an  Alfv\'{e}n wave, traveling at the local 
Alfv\'{e}n speed $v_{\rm A}(r) =  v_{\rm A}(R) R/r$, to cross the radius of the star.
If we define $t_{\rm A} \equiv R/v_{\rm A}$, then in the model analyzed here 
the Alfv\'en wave crossing time is exactly $t_{\rm A}/2 = 1$ in nondimensional units.

The scaling behavior of our solution, revealed by our nondimensional formalism,
exhibits several significant features. We find that the values of the {\it amplitudes} 
of all evolved quantities shown in Figs.~1 --3 are entirely independent 
of the magnitude of the
initial radial seed field $B_0$ given by equation (\ref{six}). Only the 
{\it timescale} of the evolution 
depends on the strength of the seed field, and it is proportional 
to the Alfv\'{e}n time associated
with that field. Specifically, no matter what the strength of the initial radial field, 
the azimuthal magnetic field will grow to the same high value 
sufficient to brake the differential
motion and drive the star to oscillate about the state of uniform 
rotation. In this state, the energy in the
azimuthal field will always grow to make up the difference between the 
rotational energy in the initial, differentially rotating configuration 
and the uniformly rotating configuration, independent of the strength of the
seed field. By the same token, the evolution timescale is independent of
the degree of differential rotation or the rotation period.

%%%%%%%%%%%%%%%%
%\vspace{-4cm}
%figure 4
\begin{inlinefigure}
\vspace{-1.7cm}
\centerline{\includegraphics[width=1.0\linewidth]{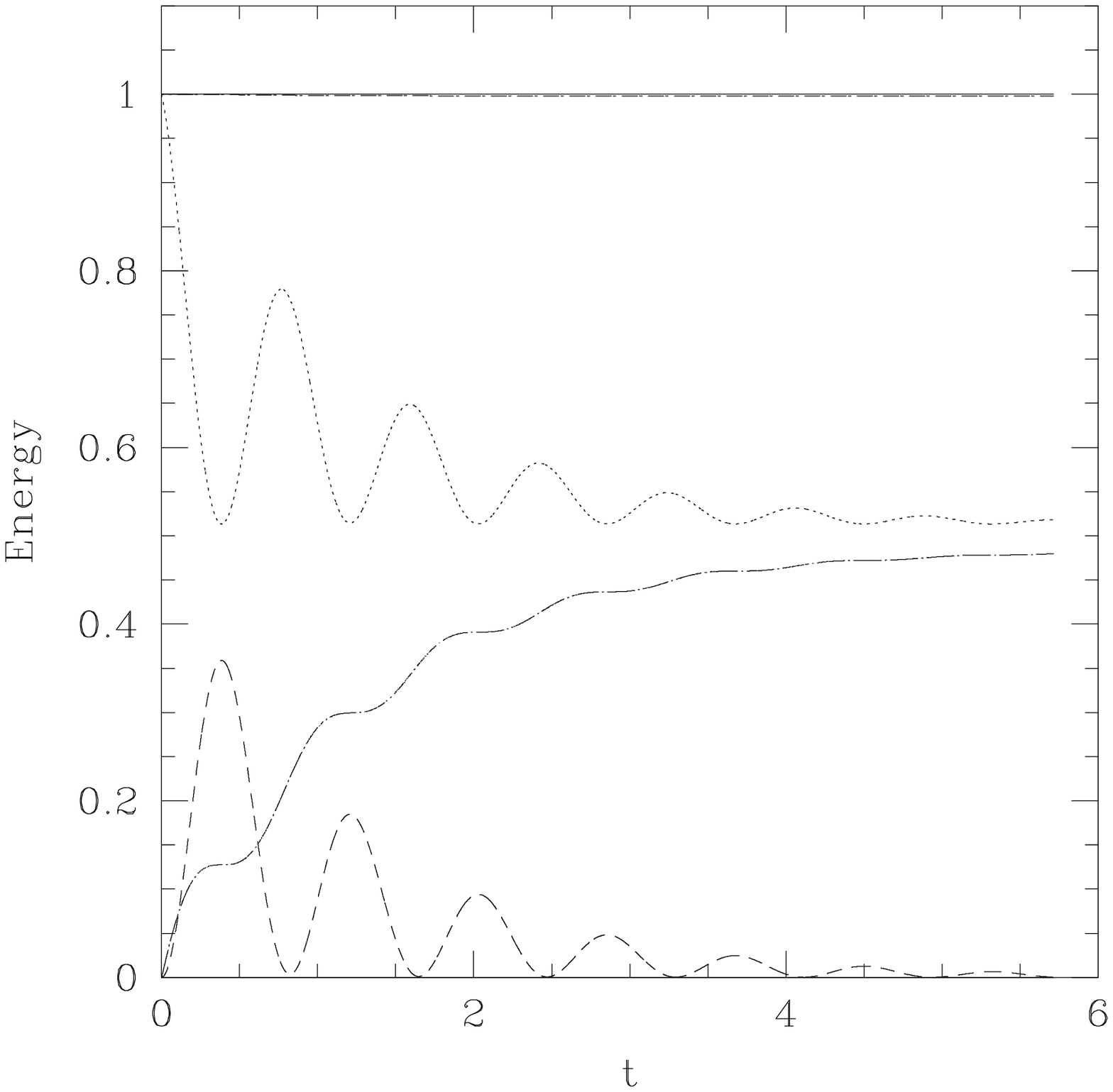}}
\vspace{-2cm}
\figcaption{Energy evolution for a differentially rotating star with internal 
viscosity $\nu = 0.2$ and
a vacuum exterior. The dotted
line shows $E_{\rm rot}$, the dashed line shows $E_{\rm mag}$, and the long-dash-dotted
line shows $E_{\rm vis}$. All energies are normalized
to $E_{\rm rot}(0)$. The sum of the energies, plotted as a short-dash-dotted line,
is conserved and remains equal to the initial 
rotational energy $E_{\rm rot}(0)$, plotted
as the solid horizontal line at $\rm{Energy} = 1$. Time is in 
nondimensional units.}
%\label{fig:analytics}}
\end{inlinefigure}
%%%%%%%%%%%%%%%%

\vspace{-0.5cm}
\subsection{Case B: ~~Nonzero viscosity with vacuum exterior}
\vspace{0.3cm}

We now consider the effect of viscous dissipation by setting 
$\nu = 0.2$ in equation (\ref{twsix}). This intermediate value is chosen in order that
viscosity in the star be sufficiently large for
dissipation to become evident in a few Alfv\'{e}n timescales, but sufficiently
small so that viscous damping of differential rotation does not completely
suppress the growth of the azimuthal magnetic field. We solve the coupled evolution
equations (\ref{twsix}) and (\ref{twseven}), subject to the same
vacuum exterior boundary and initial conditions used in Case A:
equations (\ref{ten}) -- (\ref{thirteen}) for $B_\phi$, together with 
(\ref{thone}) and (\ref{thtwo})) 
for $\Omega$. We solve these equations numerically by finite-differencing (see Appendix C).
To better treat the multiple timescales (e.g., Alfv\'{e}n {\it vs.} viscous) and
to follow the long-term damping, we employ an implicit scheme and thereby eliminate a 
Courant stability bound on our numerical timestep. We test our numerical integrations
by reproducing the analytic solution derived in Case A for $\nu = 0$ and by 
checking that energy and angular momentum are conserved 
according to equations (\ref{nineteen})
and (\ref{twtwo}).

The results of our numerical integrations are summarized in Figs.~4 - 6. In
Fig.~4 we track the evolution of the various energies with time. We observe how
the oscillations of the rotational kinetic and azimuthal magnetic field energies are now
damped by viscosity. From the evolution equation (\ref{twsix}), we may define the characteristic 
viscous dissipation timescale according to 
$t_{\nu} ={\nu}^{-1}$ in nondimensional units, which is equivalent to
$t_{\nu} = R^2/(8 \nu)$ in physical units (see equation (\ref{twfive})). For the
example considered here, the ratio of viscous to Alfv\'{e}n time is 
$t_{\nu}/t_{\rm A} = 1/(2 {\nu}) = 2.5$. The numerical results are consistent with this ratio.
%%%%%%%%%%%%%%%%
%figure 5
\begin{inlinefigure}
\vspace{-1.7cm}
\centerline{\includegraphics[width=1.0\linewidth]{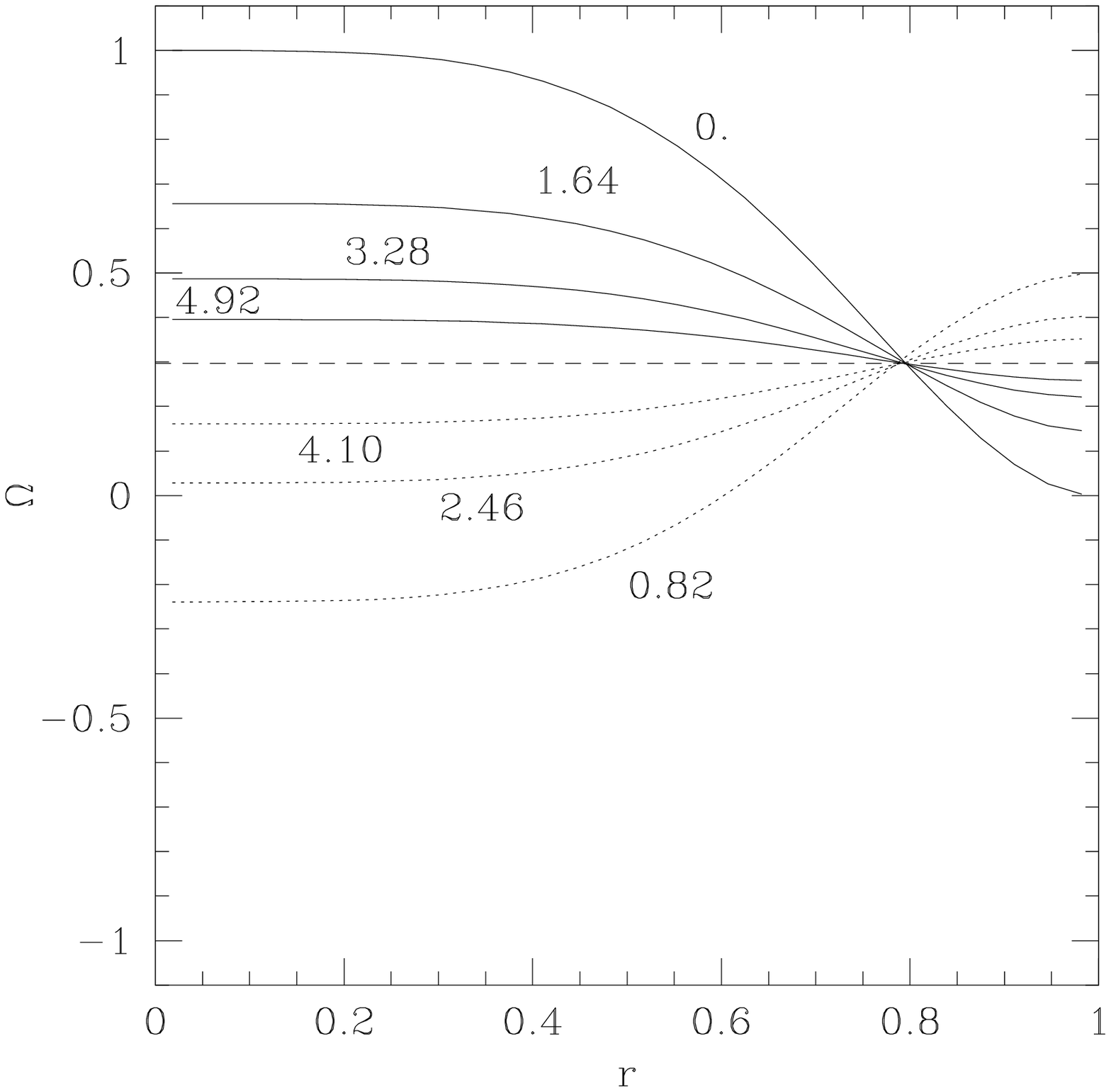}}
\vspace{-1.7cm}
\figcaption{The angular velocity profile at critical phases during the first three 
oscillation cycles for
the differentially rotating star described in Fig.~4. Curves are labeled by the
value of time. The  wave pattern undergoes damped oscillations with a period $ P = 1.64$. 
All quantities
are plotted in nondimensional units.}
%\label{fig:analytics}}
\end{inlinefigure}
%%%%%%%%%%%%%%%%
%\vspace{-1cm}
As time progresses, the rotation becomes uniform (see Fig.~5), the azimuthal
field decays back to zero (see Fig.~6), and the rotational energy difference bet
ween
the initial differentially rotating and the final uniformly rotating star
is dissipated as internal heat by the viscosity.
Angular momentum is again conserved in the star, as verified by the 
numerical results.
Fig.~4 shows that at late times the final rotational energy
approaches
$E_{\rm rot}/E_{\rm rot}(0) = 0.513$, the value already calculated in Case A for
 uniform rotation
for the given (conserved) angular momentum of our star. At the same time, the en
ergy in
viscous dissipation increases to
$E_{\rm vis}/E_{\rm rot}(0) = 1 - E_{\rm rot}/E_{\rm rot}(0) = 0.487$. Similarly
, Fig.~5 shows that
$\Omega$ undergoes
damped oscillations about the asymptotic, uniform value $\Omega_{\rm unif}$ give
n
by equation (\ref{theight}).

The asymptotic stationary state of the star, which consists of
uniform rotation, zero azimuthal magnetic field, and appreciable internal heat content
is uniquely determined by conservation of total energy and angular momentum: it 
does not depend on the strength of the assumed viscosity, only that some viscosity is
present, however small. The magnitude of the viscosity does determine the damping timescale
for differential rotation and Alfv\'{e}n wave oscillations. For a sufficiently 
high viscosity $\nu \gg 1$, differential rotation is damped  without driving
Alfv\'{e}n waves throughout the star. However, the final state is unchanged.

%%%%%%%%%%%%%%%%
\vspace{-2cm}
%figure 6
\begin{inlinefigure}
\centerline{\includegraphics[width=1.0\linewidth]{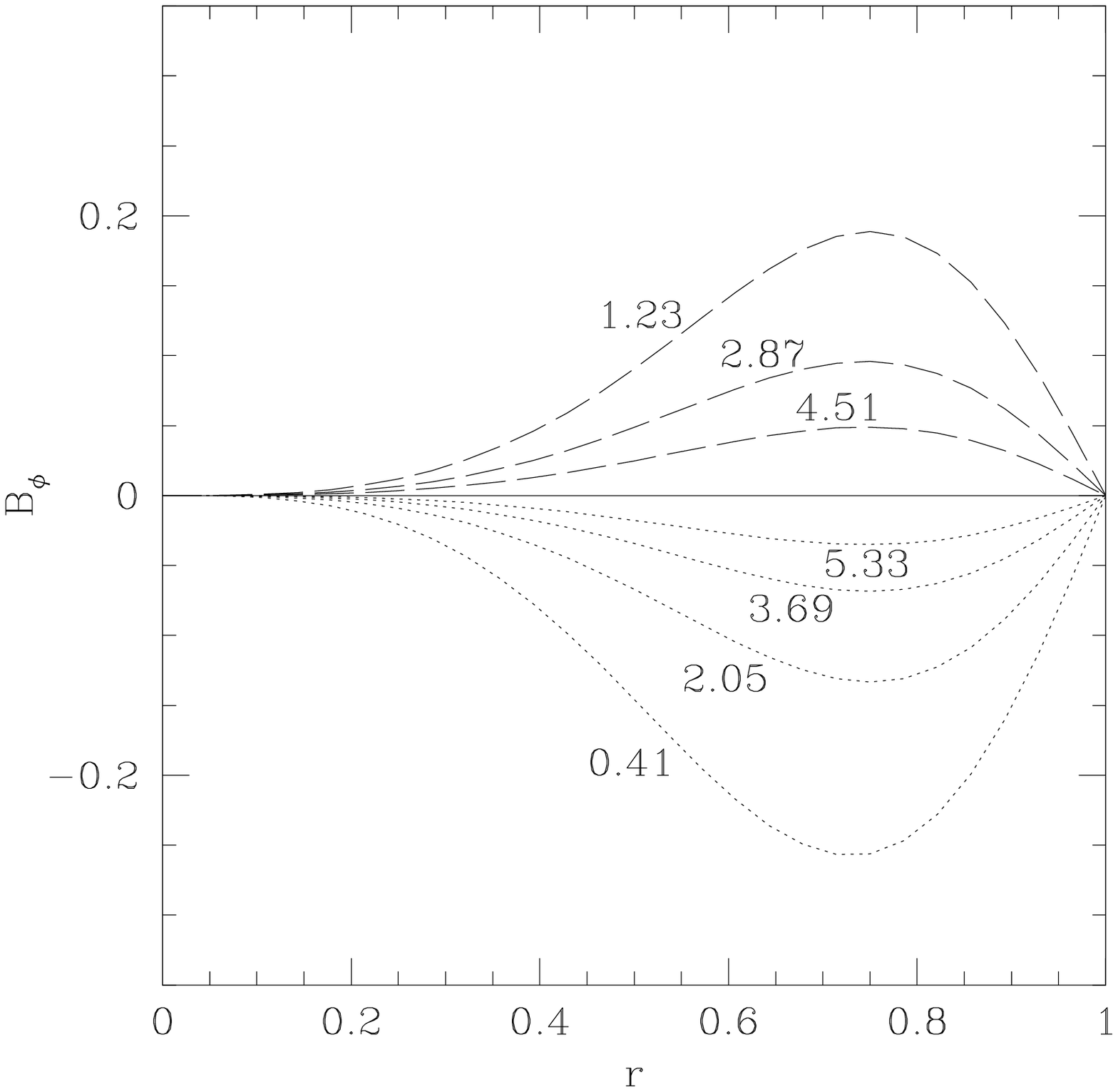}}
\vspace{-2cm}
\figcaption{The azimuthal magnetic field profile at critical phases during the first three 
oscillation cycles for
the differentially rotating star described in Fig.~4. Curves are labeled by the
value of time. The Alfv\'{e}n wave pattern undergoes damped oscillations
with a period $ P = 1.64$.
All quantities are plotted in nondimensional units.}
%\label{fig:analytics}}
\end{inlinefigure}
%%%%%%%%%%%%%%%%

\vspace{0.5cm}
\subsection{Case C: ~~$\nu = 0$ with plasma exterior}

Here we treat a differentially rotating star that has no viscosity but
is surrounded by an infinite, homogeneous medium of perfectly conducting
plasma.  We solve the coupled evolution equations (\ref{twsix}) and (\ref{twseven})
for the same boundary and initial conditions as in Case A, except at the surface,
where we now use equation (\ref{fourteen}) in place of (\ref{thirteen}) to account
for the partial transmission and reflection of the Alfv\'{e}n wave at the
star -- plasma interface. We set $\rho_{\rm ex}/\rho = 0.2$.  As in Case B, we
again solve the equations by finite differencing (see Appendix C).

Our results are summarized in Figs.~7 -- 10, where we see how the loss of
energy and angular momentum to the exterior via the transmission of
toroidal Alfv\'{e}n waves drives the star to a slower, rigidly rotating state.
In Fig.~7 we observe how the oscillatory exchange of energy between rotation and
toroidal magnetic field is damped by the outgoing Poynting energy flux at the
surface. Fig.~8 shows how the torque exerted on the star by 
the magnetic field at the surface causes the star to lose angular momentum to
the exterior plasma. Fig.~9 shows that
the angular velocity profile becomes increasingly uniform with time and that
the star ends up spinning in the {\it opposite} sense from its rotation 
at $t = 0$. By the time the star settles into uniform rotation, 
the interior azimuthal field dies away, as shown 
in Fig.~10. Figs.~7 and 8 confirm that total
energy and angular momentum are conserved, in compliance with equations (\ref{nineteen})
and (\ref{twtwo}).

%%%%%%%%%%%%%%%%
%figure 7
\begin{inlinefigure}
\vspace{-1.2cm}
\centerline{\includegraphics[width=1.0\linewidth]{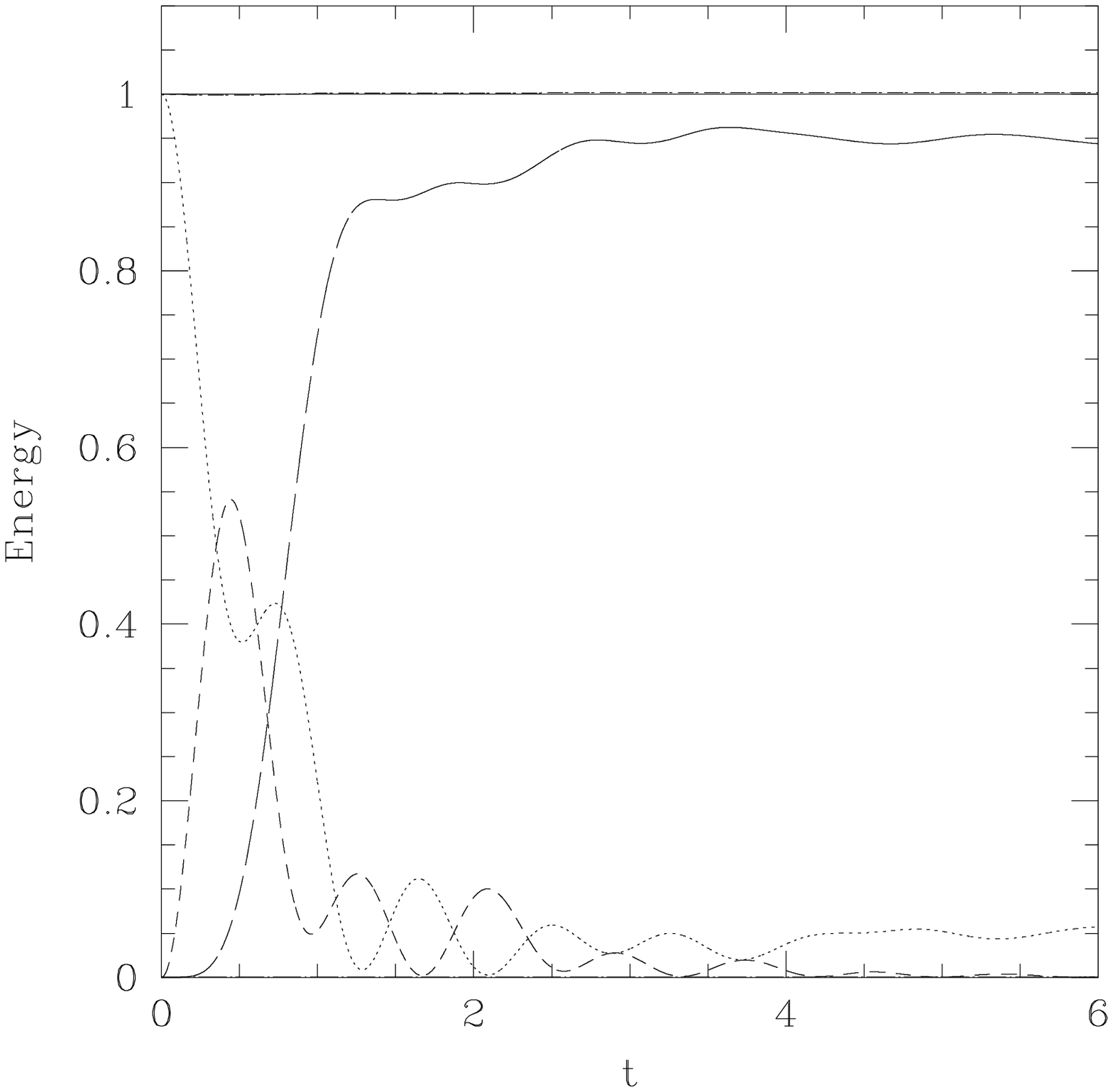}}
\vspace{-1.8cm}
\figcaption{Energy evolution for a differentially rotating, zero-viscosity star embedded
in an ambient plasma where $\rho_{\rm ex}/\rho = 0.2$. 
The dotted line shows $E_{\rm rot}$, the short-dashed line shows $E_{\rm mag}$,
and the long-dashed line shows the outgoing integrated 
Poynting energy flux $E_{\rm Poyn}$.
All energies are normalized
to $E_{\rm rot}(0)$. The sum of the energies, indicated by the dot-dashed line,
is conserved and remains
equal to the initial rotational energy $E_{\rm rot}(0)$, 
plotted as the solid horizontal line at Energy $= 1$. Time is in 
nondimensional
units.}
%\label{fig:analytics}}
\end{inlinefigure}
%%%%%%%%%%%%%%%%
%%%%%%%%%%%%%%%%
%figure 8
\begin{inlinefigure}
\vspace{-1.2cm}
\centerline{\includegraphics[width=1.0\linewidth]{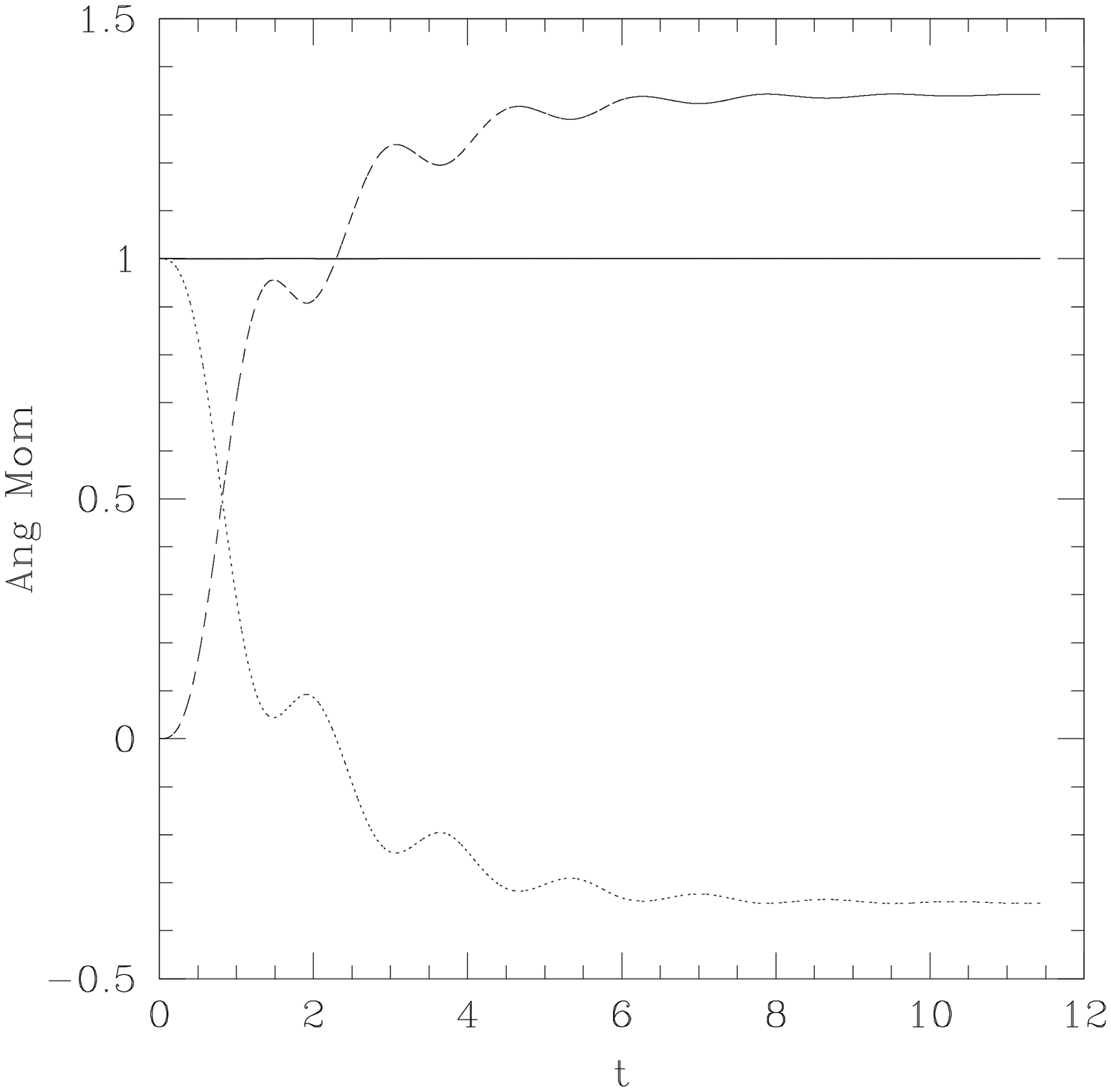}}
\vspace{-1.8cm}
\figcaption{Angular momentum evolution for the differentially rotating star depicted
in Fig.~7. The dotted curve shows the angular momentum of the star,
while the dashed curve shows the integrated torque exerted by the Maxwell
stress at the stellar surface. All contributions are normalized to
$J_{\rm rot}(0)$. The sum of all angular momenta, a dot-dashed line,
is conserved and remains equal to $J_{\rm rot}$, plotted as the solid 
horizontal line at Ang Mom $= 1$.}
%\label{fig:analytics}}
\end{inlinefigure}
%%%%%%%%%%%%%%%%

The existence of an ambient plasma is sufficient to drive the spin of the
star to low values; internal dissipation is not necessary. In this model,
the magnitude of the final (retrograde) spin of the star, $\Omega \approx -0.1$,
is insensitive to the ratio of the densities of the homogeneous ambient 
plasma to the homogeneous stellar interior, $\rho_{\rm ex}/\rho$. This ratio
influences the timescale of rotation damping only. A crude estimate suggests
that the net damping timescale is roughly
$t_{\rm A,ex} \approx t_{\rm A} (\rho/\rho_{\rm ex})^{1/2}$, for small 
values of $\rho_{\rm ex}/\rho$. This estimate accounts
for the fact that only a fraction  
$\approx 4(\rho_{\rm ex}/\rho)^{1/2} $ of the outgoing Poynting energy flux carried
by the Alfv\'{e}n wave actually 
escapes from the stellar surface into the ambient gas 
during each oscillation cycle, while the 
rest is reflected back inward (see equation [\ref{Afour}]).

%%%%%%%%%%%%%%%%
%figure 9
\begin{inlinefigure}
\vspace{-1.8cm}
\centerline{\includegraphics[width=1.0\linewidth]{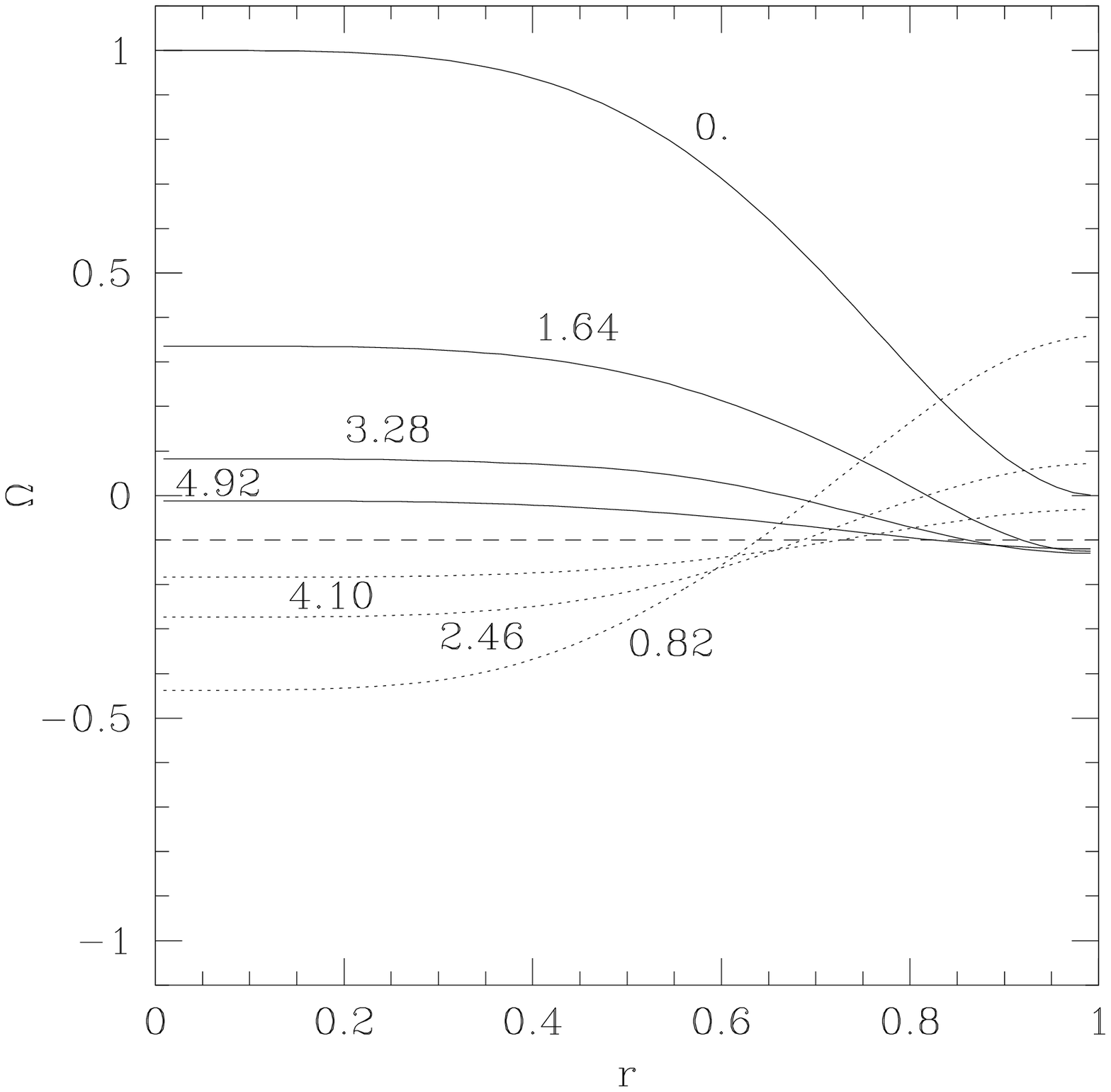}}
\vspace{-2cm}
\figcaption{The angular velocity profile at critical phases during the first three 
oscillation cycles for
the differentially rotating star described in Fig.~7. 
Curves are labeled by the
value of time. The  wave pattern undergoes damped oscillations with a 
period $ P = 1.64$. The star settles into retrograde motion with 
angular velocity $\Omega \approx -0.1$, indicated by the dashed line.
All quantities
are plotted in nondimensional units.}
%\label{fig:analytics}}
\end{inlinefigure}
%%%%%%%%%%%%%%%%
%%%%%%%%%%%%%%%%
%figure 10
\begin{inlinefigure}
\vspace{-1.8cm}
\centerline{\includegraphics[width=1.0\linewidth]{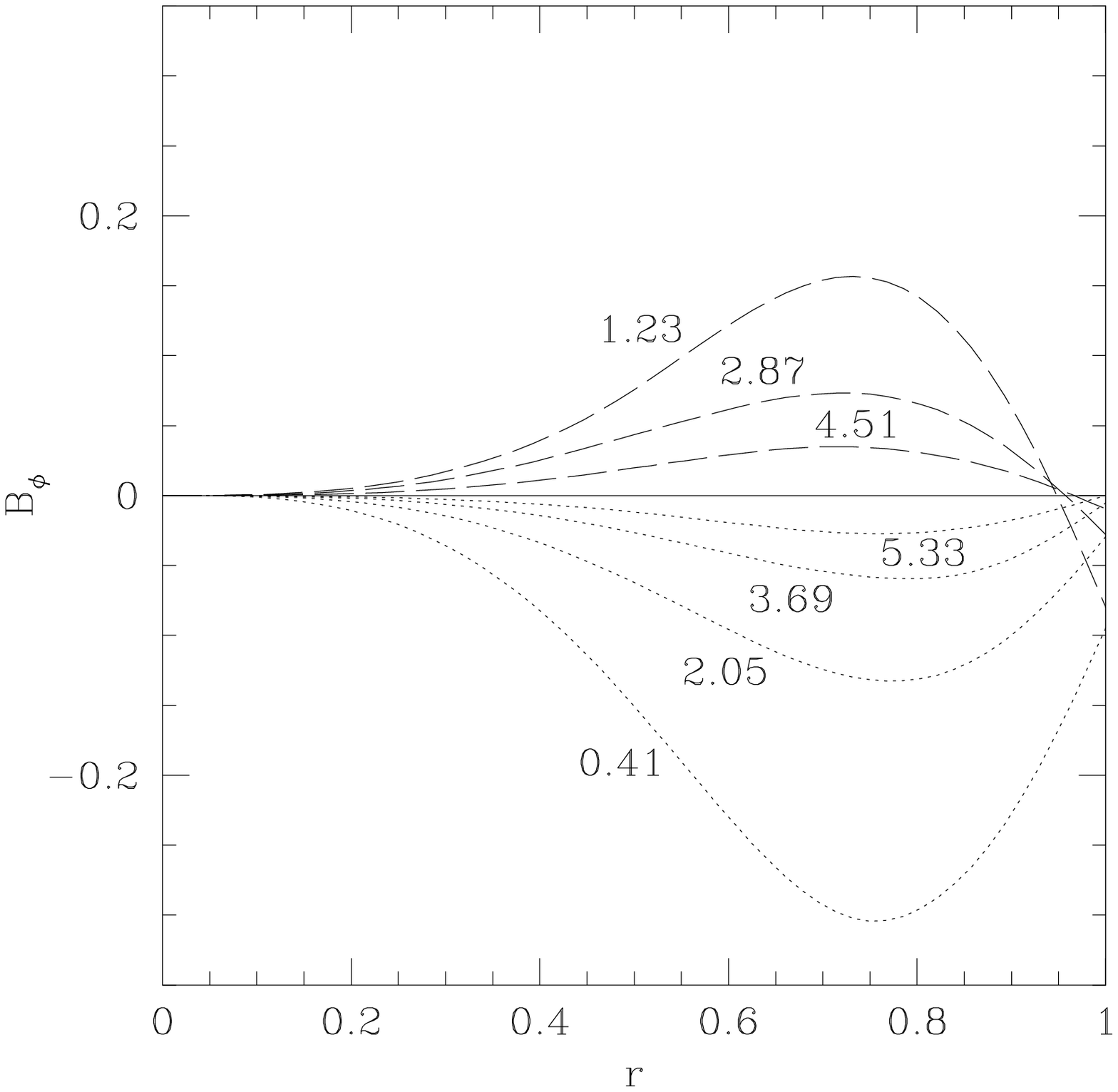}}
\vspace{-2cm}
\figcaption{The azimuthal magnetic field profile at critical phases during the first three
oscillation cycles for
the differentially rotating star described in Fig.~7. Curves are labeled by the
value of time. The Alfv\'{e}n wave pattern undergoes damped oscillations
with a period $ P = 1.64$.
All quantities are plotted in nondimensional units.}
%\label{fig:analytics}}
\end{inlinefigure}
%%%%%%%%%%%%%%%%

\section{Discussion}

The presence of a seed poloidal magnetic field in a 
differentially rotating star may be sufficient to drive the star to 
a state of uniform rotation. Differential rotation generates toroidal 
Alfv\'{e}n waves through the star and 
these waves can convert an appreciable fraction of the kinetic energy
in differential motion into magnetic field energy.
Even in the absence of internal dissipation or Poynting energy loss
to an external plasma, the conversion of rotational to magnetic energy
by these waves can remove a portion of the star's rotational support 
against gravity, which can upset the equilibrium balance in 
a stationary configuration.  In the case of
a hypermassive neutron star supported against collapse by differential
rotation, the generation of Alfv\'{e}n waves can lead to 
catastrophic collapse to a black hole, possibly accompanied by
mass loss. Such an outcome may characterize the final fate of 
neutron stars formed in binary mergers (BSS).

Our model calculations help to identify 
at least five distinct timescales that govern the evolution of a
a differentially rotating, equilibrium neutron star containing a
seed poloidal magnetic field. We evaluate them here for neutron star 
parameters consistent
with the stable remnant found in the relativisitic simulations of binary 
mergers by Shibata and Uryu (2000). These parameters are 
comparable to the values of the hypermassive, 
differentially rotating equilibrium model constructed and evolved by 
BSS, who found that their configuration was also dynamically stable.
All of these models employ a polytropic equation of state with polytropic
index $n=1$, for which there is scale freedom to choose the physical mass.
If we take the mass to be
$M \approx 2 \times 1.5M\odot \approx 3 M\odot$, a reasonable value for
a binary remnant, then 
the equatorial radius of
the star is $R \approx 5M \approx 20 {\rm km}$ and the central angular velocity
is $\Omega_0 \approx 0.3/M \approx 2 \times 10^4 {\rm Hz}$, a factor
of three times higher than the angular velocity at the equator.

The dynamical timescale associated with the star is given by
\begin{equation}
t_{\rm dyn} = \left( {R^3 \over M} \right)^{1/2}
\approx 0.15 \left({R \over 20 \mbox{km}}\right)^{3/2} 
\left({M \over 3 M_{\odot}}\right)^{-1/2} \mbox{ms}, 
\label{thnine}
\end{equation}
the time it takes the binary remnant to achieve equilibrium 
following coalescence.
This is also the time that it takes the star, once it is 
driven out--of--equilibrium 
by magnetic braking, to undergo collapse. The central rotation period 
of the remnant is
\begin{equation}
t_{\rm rot} = {2\pi \over \Omega_0} 
\approx 0.3 \left({R \over 20 \mbox{km}}\right)^{3/2} 
\left({M \over 3 M_{\odot}}\right)^{-1/2} \mbox{ms}, 
\label{for}
\end{equation}
while the period at the equator is about three times longer. The 
timescale for magnetic braking of differential rotation by
Alfv\'{e}n waves is given by  
\begin{equation}
t_ {\rm A} = {R \over v_{\rm A}}\approx 10^2 \left({B_0 \over 10^{12} \mbox{G}}\right)^{-1}
\left({R \over 20 \mbox{km}}\right)^{-1/2}
\left({M \over 3 M_{\odot}}\right)^{1/2} \mbox{s}.
\label{forone}
\end{equation}
On this timescale the angular velocity profile in the star is
signifcantly altered, and this change can drive the star far out of 
equilibrium whenever the initial rotation supplies an 
appreciable fraction of the
support against gravity. In the case of a hypermassive neutron star, which
depends on rapid differential rotation
for hydrostatic support, magnetic braking can lead to
catastrophic collapse to a Kerr black hole, possibly
accompanied by some mass and field ejection. 
A hypermassive star is a likely outcome of binary neutron star
coalescence. The resulting delayed collapse following merger
may generate a brief secondary burst of gravitational waves lasting a
time $t_{\rm dyn}$. If the time of the 
coalescence can be inferred
from the initial wave burst signal, then the measurement of this delay
should provide an estimate of the strength of the poloidal magnetic field
in the interior of the merged neutron star (BSS).

Neutron stars below the supramassive limit do not require differential
rotation for hydrostatic support. Such stars have masses at most $~ 20\%$
larger than the nonrotating, spherical TOV limit 
(see, e.g, Butterworth \& Ipser (1975);
Friedman, Ipser \& Parker (1986); Komatsu, Eriguchi and Hachisu (1989);
Cook, Shapiro \& Teukolsky 1992,1994; Salgado et al. 1994). 
\footnote{The most recent equations of state
place the maximum gravitational mass of a spherical star in the 
range $1.8 - 2.2 M_{\odot}$ (Akmal, Pandharipande
\& Ravenhall 1998), but observations of SN 1987A suggest that
it could even be as low as $1.5 M_{\odot}$ (Brown \& Bethe 1994)}
If such a star is formed with differential rotation, magnetic braking
will still drive Alfv\'{e}n waves in the star and alter the rotation profile 
on the Alfv\'{e}n time. While the star may undergo dynamical oscillations,  
it will not collapse.
Instead, viscous dissipation ultimately will drive the star to a new, 
uniformly rotating equilibrium state on a timescale
\begin{eqnarray}
t_{\nu} &\!\! = \!\! & {R^2 \over 8 \nu} \\
& \!\! \approx \!\! &  2\times 10^9 \left({R \over 
20 \mbox{km}}\right)^{23/4}\left({T \over 10^9 {\mbox K}}\right)^2
\left({M \over 3 M_{\odot}}\right)^{-5/4} \mbox{s}  ,  \nonumber 
\label{fortwo}
\end{eqnarray}
where $\nu = 347 \rho^{9/4} T^{-2} {\rm cm^2 s^{-1}}$ (Cutler \& Lindblom 1987).

If a differentially rotating neutron star is immersed in an ambient plasma
at formation, Alfv\'{e}n waves generated in the interior will propagate into 
the external medium, carrying off energy and angular momentum. According to
equation (\ref{Afour}), everytime a
wave propagates to the stellar surface it will transmit a fraction
$\approx 4 (\rho_{\rm ex} / \rho)^{1/2}, ~~ (\rho_{\rm ex} / \rho) \ll 1$,  
of its energy to the surrounding plasma, causing the star to spindown in 
an e-folding time 
\begin{eqnarray}
t_ {\rm A,ex} & \!\!=\!\! & {\mbox {ln 2} \over 4}
\left({\rho \over \rho_{\rm ex}}\right)^{1/2} t_{\rm A}  \\ 
&  \!\! \approx \!\! & 20 \left({B_0 \over 10^{12} \mbox{G}}\right)^{-1} 
\left({R \over 20 \mbox{km}}\right)^{-1/2}
\left({M \over 3 M_{\odot}}\right)^{1/2} 
\left({\rho \over \rho_{\rm ex}}\right)^{1/2} 
\mbox{s}.  \nonumber
\label{forthree}
\end{eqnarray}
({\it cf.} Spitzer 1978, eqn 13.54 and discussion therein for an alternative
argument leading to an analogous result).
Determining the final spin rate and fate of a newly formed neutron star
may thus depend on the nature of its immediate surroundings.

Our naive model calculations do not account for convective
instabilities (Pons et al. 1999), which can drive the seed magnetic
field to high values greatly 
exceeding $10^{12}{\rm G}$ (Duncan \& Thompson 1992), or other 
possible MHD instabilities,
which may also contribute to the redistribution of angular momentum
(Balbus \& Hawley 1994; Spruit 1999b). 
Our main goal was to show that even in the
simplest laminar description,
a differentially rotating
neutron star is a transient phenomenon,  
since magnetic braking and viscosity inevitably will bring a 
differentially rotating star into uniform rotation (although
not before many dynamical timescales have elapsed, according to
our timescale estimates above).  
Thus all radio pulsars are likely to be uniformly rotating. 
However, differential rotation may
characterize a nascent neutron star formed in a supernova,
following fallback, or in a merged binary. The braking of this
motion may have important consequences for gravitational wave
signals and gamma-ray bursts.  

More realistic evolutionary calculations of magnetic
braking in neutron stars should clarify some of the above issues.
One computational subtlety arises from
the inequality $t_{\rm dyn} \ll t_{\rm A}$, which holds for seed magnetic
fields of small or moderate strength (see equations [\ref{thnine}] 
and [\ref {forone}]). 
In this regime it may prove too taxing to a relativistic MHD code to 
evolve a differentially rotating star for the required length of time for
magnetic braking to take effect. An obvious solution would be to artificially
amplify the initial magnetic field strength, increasing the ratio 
$t_{\rm dyn}/t_{\rm A}$ to a computationally manageable size (while still keeping it 
much less than unity), and then scale the 
results for smaller ratios. However, if the energy in the seed magnetic
field is an appreciable fraction of the gravitational potential energy, care
must be taken to incorporate the magnetic field stresses when constructing the
initial hydrostatic equilibrium configuration. 
Another approach might be to use implicit
differencing in the calculation to avoid the Courant 
stability constraint on the evolution timestep. 
In addition, one might be able treat {\it part} of the
evolution in the quasistatic approximation, as in a typical stellar 
evolution code, up to the moment that stable equilibrium can no 
longer be sustained.  Specifically,  one might evolve the 
magnetic field and the angular velocity 
on the Alfv\'{e}n timescale via the MHD equations, but assume hydrostatic 
equilibrium is maintained on the dynamical timescale in order to 
determine the density and pressure profiles at each instant. 
Our idealized incompressible 
cylindrical star model, which remains
radially static, trivially conforms to this behavior and thereby
avoids the mismatch between the dynamical and Alfv\'{e}n timescales. 
(An additional motivation for treating this idealized problem was to 
highlight this feature of the analysis.)  We hope to tackle the more 
general problem in future computational investigations.

\acknowledgments

It is a pleasure to thank T. Baumgarte, C. Gammie and M. Shibata 
for helpful discussions. This work was supported by 
NSF Grants AST 96-18524 and NASA Grants NAG 5-7152 and NAG 5-8418 at 
the University of Illinois at Urbana-Champaign.  

\appendix
\section{Partially Transmitted/Reflected Wave Outer Boundary Condition}

If the surface of a star is the interface between a highly
conducting interior plasma of density $\rho$ and a highly conducting
exterior plasma of density $\rho_{\rm ex}$, an  
Alfv\'{e}n wave generated in the interior progating to the surface
suffers partial reflection and transmission (Roberts 1967). 
The reflection and transmission coefficients for the transverse magnetic field
amplitude are
\begin{equation}
{\cal R}\ =\ {(\rho_{\rm ex}/\rho)^{1/2}-1 \over (\rho_{\rm ex}/\rho)^{1/2}+1}
\label{Aone}
\end{equation}
and 
\begin{equation}
{\cal T}\ =\ {2(\rho_{\rm ex}/\rho)^{1/2} \over (\rho_{\rm ex}/\rho)^{1/2}+1}
\label{Atwo}
\end{equation}
The relection and transmission coefficients for the Poynting energy
flux are
\begin{equation}
{\cal R}_{\rm Poyn}\ =\ \left( {(\rho_{\rm ex}/\rho)^{1/2}-1 
\over (\rho_{\rm ex}/\rho)^{1/2}+1}\right)^2
\label{Athree}
\end{equation}
and 
\begin{equation}
{\cal T}_{\rm Poyn}\ =\ {4(\rho_{\rm ex}/ \rho)^{1/2} \over 
\left( (\rho_{\rm ex}/\rho)^{1/2}+1\right)^2}.
\label{Afour}
\end{equation}
Note that ${\cal R}_{\rm Poyn} + {\cal T}_{\rm Poyn} = 1$, in compliance with
energy conservation.

We seek to impose an approximate boundary condition on the transverse magnetic field 
just inside the surface. Decomposing the wave amplitude of the field
into its outgoing incident and ingoing reflected components, we may 
write
\begin{equation}
B_{\phi}(t,r) = f(t-r/v_{\rm A})+{\cal R}f(t+r/v_{\rm A})
\label{Afive}
\end{equation}
where $v_{\rm A}$ is the 
Alfv\'{e}n propagation speed at the surface. In writing 
equation (\ref{Afive}) we 
approximate the wave to be planar, which is strictly true only
for wavelengths $\lambda_{\rm A} \ll R$. Differentiating equation (\ref{Afive})
yields
\begin{equation}
f' = {{\partial B}_{\phi} \over {\partial t}}{1 \over (1+ {\cal R})}
= -{{\partial B}_{\phi} \over {\partial r}}{v_{\rm A} \over (1-{\cal R})}
\label{Asix}
\end{equation}
where the prime denotes a total derivative. We may then recast 
equation (\ref{Asix}) as
\begin{equation}
{{\partial B}_{\phi}(t,R) \over {\partial t}} +
{{\partial B}_{\phi}(t,R) \over {\partial r}}{v_{\rm A}}
{(1+{\cal R}) \over (1 - {\cal R})} = 0, 
\label{Aseven}
\end{equation}
which provides the desired boundary condition.

\section{Uniform Rotation as the Lowest Energy Configuration }

Here we prove that for fixed angular momentum, the angular velocity
profile that gives a stationary
star of lowest energy is one of uniform rotation. 
To find the lowest energy state, set the azimuthal magnetic field energy to
zero and take the total energy of the 
configuration to be rotational kinetic energy,
\begin{equation}
E= \int^1_0 dr^2 E(\Omega, r^2) =
\int^1_0 dr^2 \Omega^2 r^2,
\label{Bone}
\end{equation}
The conserved angular momentum is given by
\begin{equation}
J = \int^1_0 dr^2 J(\Omega,r^2)= \int^1_0 dr^2 \Omega r^2.
\label{Btwo}
\end{equation}
The lowest energy configuration is found by varying the energy 
subject to the constraint of fixed angular momentum, i.e.,
by varying the integral 
\begin{equation}
{\cal I}= E +{\lambda}J
\label{Bthree}
\end{equation}
where $\lambda$ is a (constant) Lagrange multiplier. 
Setting the variation equal
to zero yields the Euler-Lagrange equation for the functional
\begin{equation}
{\cal L}(\Omega,\Omega',r^2)= E(\Omega,r^2) +{\lambda}J(\Omega,r^2).
\label{Bfour}
\end{equation}
The Euler-Lagrange equation gives
\begin{equation}
{d{\cal L} \over d {\Omega} }= (2 {\Omega} + {\lambda})r^2 = 0,
\label{Bfive}
\end{equation}
or
\begin{equation}
{\Omega}=-{{\lambda} \over 2} = {\rm constant},
\label{Bsix}
\end{equation}
which is the desired result. 

\section{Finite Difference Equations}

To finite difference the coupled evolution equations (\ref{twsix}) and
(\ref{twseven}) we introduce a uniform radial grid ${r_i},~i=1,2,\ldots,
i_{max}$, which extends from $r_1=0$ to $r_{i_{max}}=1$. The angular
velocity $\Omega$ is defined at the midpoints of the 
radial zones while the magnetic field $B$ is
defined on the zone boundaries. We adopt an implicit evolution scheme in
which the differencing is first order in time but second order in space.
Care is taken when finite differencing the radial gradients so that they
reflect the correct regularity behavior at the origin, where
$B \sim a_1r$, while $\Omega \sim a_0+a_2r^2$; here $a_0$, $a_1$ and
$a_2$ are constants.

To evolve $\Omega$ for a timestep $\Delta t = t_{n+1}-t_n$ 
we difference equation (\ref{twsix}) according to 
\begin{eqnarray}
\label{Cone}
{\Omega_{i+1/2}^{n+1}-\Omega_{i+1/2}^n \over \Delta t} 
&&=\;\;
{1 \over r_{i+1/2}^2}{ \left( r_{i+1}B_{i+1}^{n+1}-r_iB_i^{n+1} \right)
\over \left (r_{i+1}^2 - r_i^2 \right) }\\
\vspace {.5cm}
&&+\;\; {\nu \over r_{i+1}^4 - r_i^4}
\left( 
r_{i+1}^4 { \left( \Omega_{i+3/2}^{n+1}-\Omega_{i+1/2}^{n+1} \right)
\over \left( r_{i+3/2}^2-r_{i+1/2}^2 \right) }-
r_{i}^4 { \left( \Omega_{i+1/2}^{n+1}-\Omega_{i-1/2}^{n+1} \right)
\over \left( r_{i+1/2}^2-r_{i-1/2}^2 \right) } \right) \nonumber
\end{eqnarray}
while to evolve $B$ we difference equation (\ref{twseven}) according to
\begin{equation}
{B_i^{n+1}-B_i^n \over \Delta t} =
r_i { \left( \Omega_{i+1/2}^{n+1}-\Omega_{i-1/2}^{n+1} \right) \over
\left( r_{i+1/2}^2-r_{i-1/2} \right) }
\label{Ctwo}
\end{equation}
Boundary condition (\ref{twelve}) at the origin requires 
\begin{equation}
B_1^{n+1}=0. 
\label{Cthree}
\end{equation}
The outer boundary condition depends on the nature of the exterior medium:
\begin{equation}
B_{i_{max}}^{n+1}=0, ~~~~{\rm (vacuum\ exterior)}
\label{Cfour}
\end{equation}
according to equation (\ref{thirteen}),
or
\begin{equation}
B_{i_{max}}^{n+1}=B_{i_{max}}^n - {\Delta t \over 
2 \left( r_{i_{max}} - r_{i_{max}-1} \right) }\left( {1+{\cal R} \over
1-{\cal R}} \right) \left ( B_{i_{max}}^n - 
B_{i_{max}-1}^n \right), ~~~~{\rm (plasma\ exterior)},
\label{Cfive}
\end{equation}
according to equation (\ref{fourteen}).
The two prescriptions agree in the case of a vacuum exterior, where
$\rho_{\rm ex}=0$ and
${\cal R}=-1$, whereby condition (\ref{Cfive}) gives 
$B_{i_{max}}^{n+1}=B_{i_{max}}^{n}=,\dots,=B_{i_{max}}^1=0$.

Substituting finite difference equation (\ref{Ctwo}) into (\ref{Cone}) and using
equation (\ref{Cthree}) at the inner boundary and equation (\ref{Cfour})
or (\ref{Cfive}) at the outer boundary yields a single
tridiagonal matrix equation for $\Omega^{n+1}$. We invert this 
tridiagonal matrix
at each timestep, after which we use equation (\ref{Ctwo}) again 
to get $B^{n+1}$.  Since we use implicit differencing in time, the upper bound on the 
size of our integration timestep $\Delta t$ is set by accuracy and not
stability criteria.

\end{document}